\documentclass[a4paper,twocolumn,superscriptaddress]{revtex4}

\usepackage{array}
\usepackage{multirow}
\usepackage{mathtools}
\usepackage{bm,amsmath}
\usepackage{natbib}
\usepackage{graphicx}
\usepackage{xcolor}
\usepackage{hyperref}
\usepackage{soul}
\usepackage{cancel}
\hypersetup{
     colorlinks,
     linkcolor={red!90!black},
     citecolor={black!10!blue},
     urlcolor={blue!80!black}
     }

\newcolumntype{L}[1]{>{\raggedright\let\newline\\\arraybackslash\hspace{0pt}}m{#1}}
\newcolumntype{C}[1]{>{\centering\let\newline\\\arraybackslash\hspace{0pt}}m{#1}}
\newcolumntype{R}[1]{>{\raggedleft\let\newline\\\arraybackslash\hspace{0pt}}m{#1}}

\newcommand{\f}{\frac}

\newcommand{\Eu}{E_\uparrow}
\newcommand{\Ed}{E_\downarrow}
\newcommand{\Ou}{O_\uparrow}
\newcommand{\Od}{O_\downarrow}
\newcommand{\mC}{\mathcal{C}}
\newcommand{\M}{\mathcal{M}}

\begin{document}

\title{Size-Stretched Exponential Relaxation in a Model with Arrested States}

\author{Vaibhav Gupta}
\email{vaibhav.1395@gmail.com}
\affiliation{TIFR Centre for Interdisciplinary Sciences, Tata Institute of Fundamental Research, Gopanpally, Hyderabad-500107, India}
\affiliation{Loomis Laboratory of Physics, University of Illinois at Urbana–Champaign, Urbana, IL 61801}

\author{Saroj Kumar Nandi}
\email{saroj@tifrh.res.in}
\affiliation{TIFR Centre for Interdisciplinary Sciences, Tata Institute of Fundamental Research, Gopanpally, Hyderabad-500107, India}

\author{Mustansir Barma}
\email{barma@tifrh.res.in}
\affiliation{TIFR Centre for Interdisciplinary Sciences, Tata Institute of Fundamental Research, Gopanpally, Hyderabad-500107, India}

\begin{abstract}
We study the effect of rapid quench to zero temperature in a model with competing interactions, evolving through conserved spin dynamics. In a certain regime of model parameters, we find that the model belongs to the broader class of kinetically constrained models, however, the dynamics is different from that of a glass. The system shows stretched exponential relaxation with the unusual feature that the relaxation time diverges as a power of the system size. Explicitly, we find that the spatial correlation function decays as $\exp(-2r/\sqrt{L})$ as a function of spatial separation $r$ in a system with $L$ sites in steady state, while the temporal auto-correlation function follows $\exp(-(t/\tau_L)^{1/2})$, where $t$ is the time and $\tau_L$ proportional to $L$. In the coarsening regime, after time $t_w$, there are two growing length scales, namely $\mathcal{L}(t_w) \sim t_w^{1/2}$ and $\mathcal{R}(t_w) \sim t_w^{1/4}$; the spatial correlation function decays as $\exp(-r/ \mathcal{R}(t_w))$. Interestingly, the stretched exponential form of the auto-correlation function of a single typical sample in steady state differs markedly from that averaged over an ensemble of initial conditions resulting from different quenches; the latter shows a slow power law decay at large times.
\end{abstract}
\maketitle

\section{Introduction}
Systems which are cooled rapidly often reach arrested states in which the kinetics is strongly constrained. Such systems then often show anomalously slow dynamics, manifested through slow decays of time-dependent correlation functions. A variety of forms is possible for such decays, but a form which is often encountered is that of a stretched exponential relaxation (SER)
\begin{equation}\label{sereq}
\phi(t)\sim \exp(-t/\tau)^\beta
\end{equation}
where $\phi(t)$ is an autocorrelation function over a time stretch $t$, the time $\tau$ sets the scale for the decay, and the stretching exponent $\beta$ satisfies $0<\beta<1$. First observed in capacitor discharges by Kohlrausch \cite{kohlrausch1854} and subsequently by Williams and Watts in the context of dielectric relaxation \cite{williams1970}, this form of decay has been found in a host of systems, some with quenched disorder and others without; representative systems are discussed in Section \ref{SERoverview}.

Our focus in this paper is on the decay of the spin autocorrelation function $\phi(t)$ in a particular type of arrested state that arises in a 1D Ising model with competing first and second neighbor interactions and conserved dynamics \cite{das1999}. 
Competing interactions arise in magnetic rare earths through the Ruderman-Kittel-Kasuya-Yosida (RKKY) interaction \cite{redner1998}. Such interactions can also be realized in artificial spin chains \cite{nguyen2017} where one can experimentally reach the regime of interest in this paper (that is weak antiferromagnetic second neighbor interaction). Competing interactions between close-by Ising spins in one dimension also arise in quite a different context, namely stacking dynamics in polytypes, where Ising pseudo-spins $+1$ and $-1$ are associated with cyclic and anti-cyclic $ABC$ stackings respectively \cite{yeomans1988}. The kinetics of stacking transitions such as $3C - 6H$ \cite{jepps1980,kabra1986} involves conservation laws in related competing-interaction spin models \cite{das1999a}.

On quenching instantaneously from infinite temperature $T$ to $T=0$, the model considered in \cite{das1999} is known to show a number of distinct types of arrested states, depending on the ratio of the coupling constants. The arrested state of interest to us here is one in which $\phi(t)$ follows Eq. (\ref{sereq}) but with the exceptional feature that the relaxation time $\tau$ depends strongly on the system size $L$
\begin{equation}\label{sser}
\tau=\tau_L\sim L^z.
\end{equation}
We refer to relaxation characterized by Eqs. (\ref{sereq}) and (\ref{sser}) as {\em Size-stretched exponential relaxation} (SSER), to emphasize that the stretching depends on system size. As discussed in Section \ref{SERoverview}, within the class of stretched exponential decays, SSER is uncommon and so far, to the best of our knowledge, only one other class of systems with this property has been identified. 

It is interesting to ask about the qualitative physical features that give rise to SSER in the arrested state under study here. During the quench to $T=0$, the system is able to anneal out most localized high energy excitations, but a dilute gas of mobile excitations (domain walls) remains, whose number $N$ varies sub-linearly $\sim L^\alpha$ ($\alpha<1$) and stays constant in time owing to a conservation law. The subsequent time evolution of the state involves microscopic moves which conserve energy and are consequently constrained by the local configurations of spins in the close vicinity. Thus the evolution rules fall within the class of Kinetically Constrained Models (KCM) \cite{ritort2003,fredrickson1985,fredrickson1988} discussed in the next section. However, unlike KCMs studied earlier, these rules produce an SSER, as embodied in Eqs. (\ref{sereq}) and (\ref{sser}), as a consequence of the number of excitations growing subextensively.

The main results of this work are as follows: (i) we show that the spatial correlation function in the steady state decays exponentially with a length scale varying as $\sqrt{L}$. (ii) For a rapid quench to zero-temperature, the coarsening length scale, $\mathcal{L}(t_w)$, at waiting time $t_w$ varies as $t_w^{1/2}$.
But there is also a second length scale $\mathcal{R}(t_w)$ which grows as $t_w^{1/4}$, and governs the decay of the two-point correlation function. (iii) The auto-correlation function in the steady state reached on starting with a typical initial condition shows a stretched-exponential relaxation with the stretching exponent $\beta=1/2$; the unusual feature is that the relaxation time depends on 
system size (SSER). When averaged over initial conditions, the auto-correlation function shows distinctly different behaviour, namely SSER followed by a power law decay.

The rest of the paper is organized as follows: A brief overview of stretched exponential decays in various systems is given in Section \ref{SERoverview}. In Section \ref{modeldetails}, we define the model along with the dynamical rules and also recall different types of arrested states which ensue on quenching the system to $T=0$ depending on parameter values. We discuss some aspects of the arrested state of interest in Sec. \ref{domainwalls}, while in Sec. \ref{results}, we present quantitative results for statics and dynamics, including the coarsening regime. 
In Sec. \ref{mappingtoSEP}, we obtain analytic results by relating our system to the simple exclusion process \cite{spitzer1970,liggettbook}, and finally conclude in Sec. \ref{disc} with a discussion of our results. 

\section{EARLIER WORK ON STRETCHED EXPONENTIAL DECAYS}
\label{SERoverview}
In this section, we give a brief overview of different classes of models which show a stretched exponential decay of correlations (Eq. (\ref{sereq})) with a view to providing a context for our results.

A well-known mechanism that produces a stretched exponential decay involves averaging over different regions of the system, with individual exponential relaxations running in parallel. A suitable distribution of relaxation times can then lead to a stretched exponential form for the overall relaxation function \cite{klafter1986,mauro2018}. In fact, as discussed in Sec. \ref{modeldetails}B below, this mechanism also leads to SER in one of the possible arrested states (not the one of primary interest in this paper) in the model under study here. The dynamic heterogeneity picture of glassy materials, as discussed below, invokes and supports this mechanism for glassy relaxation \cite{ediger2000}. In a variant, Palmer {\it et al} \cite{palmer1984} proposed a general mechanism where an individual region has an exponential relaxation, which however is hierarchically constrained. Trap models \cite{scher1973,grassberger1982,phillips1996,phillips2006,naumis2012,potuzak2011} consider the diffusion of excitations which reach randomly distributed static traps, in which case the long-time behavior is controlled by particle motion in large trap-free regions. Stretched exponential relaxation also arises in continuous time random walks with a broad distribution of pausing times \cite{shlesinger1984, klafter1986}.

 In the ordered phase of the 2D Ising model, the autocorrelation function exhibits SER arising from the relaxation of rare, long-lived droplets of the minority phase \cite{huse1987,tang1989}. Rare clusters also dominate the $T \rightarrow 0$ dynamics of a 1D Ising ferromagnet with quenched disorder in the coupling, again leading to SER \cite{dhar1980}.

SER is one of the hallmarks of glassy dynamics \cite{ediger2000,ritort2003}. Different theories, ranging from microscopic theories like the mode-coupling theory (MCT) \cite{das2004,goetzebook,reichman2005}, to phenomenological theories such as the free volume theory \cite{turnbull1961} the Adam-Gibbs-DiMarzio theory \cite{adam1965,gibbs1958} as well as the random first order transition theory \cite{lubchenko2007,xia2001,giuliorfot} have put forward mechanisms for such relaxations. Within these theories, the stretching exponent is related to the length scale of dynamic heterogeneity in the former, whereas it is related to the length-scale of spatially correlated domains in the the latter class of theories.

A class of simple models, known as Kinetically Constrained Models (KCMs) \cite{ritort2003}, have been proposed to understand some properties of a glassy system. A variety of KCMs  have been studied in the last several decades \cite{cornell1991,fredrickson1985,fredrickson1988,skinner1983}, with the constraint in the dynamics being included in different forms, for instance dynamics with local constraints such as the magnetization-conserving Kawasaki dynamics.  
In models with an infinite number of conservation laws, the configuration space divides up into an exponentially large number of sectors with different forms of slow relaxation,  including sectors which exhibit stretched exponential decays \cite{barma1994}.
Although these models show some glass-like features, the precise mechanism for SER in glassy systems remains unclear.

An interesting subclass of KCMs of particular relevance to the current work invoke the action triggered at a site by the arrival of diffusing entities, whose total number is conserved. Thus Glarum \cite{glarum1960} and Bordewijk \cite{bordewijk1975} argue that an SER with $\beta=1/2$ describes relaxation of a molecule, caused by diffusing defects of the liquid structure reaching it. In another context, Skinner \cite{skinner1983} studied a model of polymer dynamics, which was mapped onto a $1D$ Ising model. At low-$T$, only energy-conserving moves are allowed, implying a conserved number of DWs, resulting in local spin relaxation following an SER with $\beta=1/2$. This form of SER was established rigorously by Spohn \cite{spohn1989} who derived upper and lower bounds on $\tau$ in Eq. (\ref{sereq}) in a model with a conserved number of DWs, while a generalized model was considered in \cite{godreche2015}. In the arrested state studied in this paper, we will show that similar DW dynamics holds, but the result is an SSER with $\beta=1/2$ and $\tau_L\sim L$.



We close this section by alluding to the only other example of SSER that we are aware of. Systems of particles with mutual exclusion driven by a fluctuating surface are known to exhibit fluctuation-dominated phase ordering (FDPO). The autocorrelation function of particle occupancies exhibits SSER. This result is supported by an analytic calculation of $\phi(t)$ for a related coarse-grained depth model \cite{chatterjee2006} which exhibits SSER with $\beta=1/4$ and $\tau_L\sim L^2$. 


\section{Model and Arrested States}
\label{modeldetails}
\subsection{Hamiltonian and Dynamics}
We study the relaxation dynamics to, and in, an arrested state in a simple 1D model with competing interactions, namely the axial next-nearest neighbor Ising (ANNNI) model \cite{selke1988,yeomans1988}.  We consider the ANNNI model on a 1D lattice with periodic boundary conditions described by the Hamiltonian
\begin{equation}\label{hamiltonian}
\mathcal{H}=-J_{1}\sum_{i=1}^{L}S_{i}S_{i+1}+J_{2}\sum_{i=1}^{L}S_{i}S_{i+2},
\end{equation}
where $J_{1}$ and $J_{2}$ are the coupling constants corresponding to the nearest and the next-nearest neighbour interactions respectively, $S_{i}=\pm1$ is an Ising spin variable at the $i^{\text{th}}$ site and $L$ is the size of the system. 
If the next-nearest neighbour interaction is antiferromagnetic ($J_2>0$), it competes with the nearest neighbour interaction $J_1$, which could be either ferromagnetic or anti-ferromagnetic. In this study, we assume that $J_1>0$ and define $j=J_2/J_1$.
The time evolution of the system proceeds through double spin (DSF) dynamics, wherein only pairs of adjacent parallel spins are flipped: $\uparrow\, \uparrow \,\, \leftrightarrow \,\, \downarrow\,\downarrow$. 

Let us define the sub-lattice magnetizations as
\begin{align}
M_{\mathrm{even}}=\sum_{j=1}^{L/2} S_{2j};\,\,\,\, M_{\mathrm{odd}}=\sum_{j=1}^{L/2}S_{2j-1},
\end{align}
The staggered magnetization, defined as difference of the sub-lattice magnetizations, $\mathcal{M}\equiv M_{\mathrm{even}}-M_{\mathrm{odd}}$ evidently remains conserved as every DSF move changes both $M_{\mathrm{even}}$ and $M_{\mathrm{odd}}$ equally.

Under a transformation which flips every spin on the even sublattice, the nearest neighbour coupling $J_{1}$ reverses its sign but $J_{2}$ does not. The DSF dynamics then maps into spin-exchange or Kawaski dynamics $\uparrow\, \downarrow \,\, \leftrightarrow \,\, \downarrow\,\uparrow$.  In other words, the DSF dynamics with the ferromagnetic nearest-neighbor interaction and antiferromagnetic next nearest neighbor interaction becomes equivalent to the Kawasaki dynamics with both the nearest neighbor and next nearest neighbor interactions being antiferromagnetic.
The two descriptions are thus equivalent; we use DSF throughout in this paper. It is simple to transcribe results to the Kawsaki dynamics description. For instance, single-point correlation functions such as spin autocorrelation functions remain unchanged, as do correlation functions between two sites on the same sub-lattice.


\subsection{Arrested States}
Consider performing an instantaneous quench to temperature $T=0$ from an infinite-temperature random state. The $T=0$ condition only allows those DSF moves which lead to a negative or zero energy change, as computed using the Hamiltonian, Eq. (\ref{hamiltonian}). Initially the system enters a coarsening regime in which both negative-energy change and zero-energy change moves occur, and the overall energy decreases as a result. Ultimately, this ceases when the system reaches the steady state (here, the arrested state), where only the zero-energy change moves operate. The set of allowed DSF moves depends on the ratio of couplings $j$. There are five different regions along the $j$-axis, each corresponding to a type of arrested state \cite{das1999}. The local environment which affects the orientation of a pair of adjacent parallel spins consists of the nearest and next-nearest neighbors of these two spins. Since each spin can be either up or down, there is a total of 16 possible distinct local environments for a given pair of spins; these are listed in \cite{das1999}.

Each of the five arrested states has different characteristics. In particular, the arrested state which occurs when $j>1$, called the Inhomogeneous Quiescent and Active (IQA) state, was studied in detail in \cite{das1999}. In this state, active and quiescent regions of varying lengths alternate in space. Each active region relaxes exponentially, with different relaxation times for each. This leads to an SER for the auto-correlation function $\sim\exp(-(t/\tau)^\frac{1}{3})$ \cite{das1999}.

In this paper, we concentrate on a different arrested state which arises when $0\leq j<0.5$. This state has an interesting structure, with large ferromagnetically ordered domains separated by mobile domain walls, whose number is conserved.  This state with Conserved Mobile Domain Walls also shows a stretched exponential relaxation with the auto-correlation function decaying as $\sim\exp(-A(t/\tau_L)^{1/2})$  but the mechanism behind this decay is quite different from that in the IQA state. Notably, while $\tau$ in the IQA state is $L$-independent, in this state $\tau_L $ grows with system size $\sim L$, implying that the relaxation follows SSER. This is one of the main results of this paper. We now turn to a discussion of the properties and dynamics in this phase.

\begin{table*}
\centering
\begin{tabular}{|C{4cm}| C{1cm} |C{2cm} C{0.7cm} C{2cm}|C{2cm}|}
\hline
&&&&&\\
 Nature of the MC moves &  &  &  &  & $\Delta E$\tabularnewline
 \hline
 &&&&&\\
\multirow{4}{*}{Energy Lowering Moves} & (a) & $\underline{\uparrow}\:\underline{\downarrow}\:\uparrow\uparrow\:\underline{\downarrow}\:\underline{\uparrow}$ & $\longrightarrow$ & $\underline{\uparrow}\:\underline{\downarrow}\:\downarrow\downarrow\:\underline{\downarrow}\:\underline{\uparrow}$ & $-4$\tabularnewline
&&&&&\\
 & (b) & $\underline{\downarrow}\:\underline{\downarrow}\:\uparrow\uparrow\:\underline{\downarrow}\:\underline{\downarrow}$ & $\longrightarrow$ & $\underline{\downarrow}\:\underline{\downarrow}\:\downarrow\downarrow\:\underline{\downarrow}\:\underline{\downarrow}$ & $-8(0.5-j)$\tabularnewline
 &&&&&\\
 & (c) & $\underline{\uparrow}\:\underline{\uparrow}\:\uparrow\uparrow\:\underline{\downarrow}\:\underline{\uparrow}$ & $\longrightarrow$ & $\underline{\uparrow}\:\underline{\uparrow}\:\downarrow\downarrow\:\underline{\downarrow}\:\underline{\uparrow}$ & $-4j$\tabularnewline
 &&&&&\\
 & (d) & $\underline{\uparrow}\:\underline{\downarrow}\:\uparrow\uparrow\:\underline{\downarrow}\:\underline{\downarrow}$ & $\longrightarrow$ & $\underline{\uparrow}\:\underline{\downarrow}\:\downarrow\downarrow\:\underline{\downarrow}\:\underline{\downarrow}$ & $-4(1-j)$\tabularnewline
 &  &  &  &  & \tabularnewline
 \hline
 &&&&&\\
\multirow{2}{*}{Zero Energy Cost Moves} & (e) & $\underline{\uparrow}\:\underline{\uparrow}\:\uparrow\uparrow\:\underline{\downarrow}\:\underline{\downarrow}$ & $\longleftrightarrow$ & $\underline{\uparrow}\:\underline{\uparrow}\:\downarrow\downarrow\:\underline{\downarrow}\:\underline{\downarrow}$ & $0$\tabularnewline
&&&&&\\
 & (f) & $\underline{\uparrow}\:\underline{\downarrow}\:\uparrow\uparrow\:\underline{\uparrow}\:\underline{\downarrow}$ & $\longleftrightarrow$ & $\underline{\uparrow}\:\underline{\downarrow}\:\downarrow\downarrow\:\underline{\uparrow}\:\underline{\downarrow}$ & $0$ \tabularnewline
 &&&&&\\
 \hline
\end{tabular}
\caption{Allowed moves (those that do not raise the energy) in the double spin flip dynamics for a quench to $T=0$ from a high temperature random initial configuration when $0 \leq j<0.5$.}
\label{table:dsf_moves}
\end{table*}

\section{Domain Wall Description}
\label{domainwalls}
Consider a quench from a completely disordered state to $T=0$, in the coupling-constant range $0 \le j < 0.5$.  The allowed moves are shown in Table 1, where energy-lowering moves have been shown separately from equal-energy moves. The former lead to the approach to steady state, while the latter operate even in the arrested steady state, where they lead to nontrivial decays of spin autocorrelation functions.

\begin{figure}[h]
	\includegraphics[width=8.6cm]{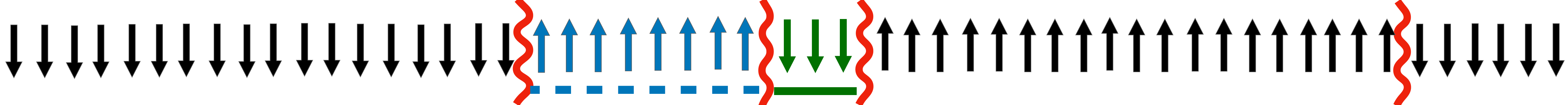}
	\caption{An arbitrary spin configuration. The domains with even number of spins (E) ultimately get annihilated in the DSF dynamics whereas those with odd number of spins (O) cannot be annihilated. The curly lines depict domain walls, and an example of an E domain (eight blue spins, underlined with dots) and an O domain (three green spins with a solid underline) are shown.}
	\label{spin_config_schematic}
\end{figure}

Domain walls (DWs) separate successive parallel-spin segments and hence lie between every pair of adjacent parallel spin segments (Fig. \ref{spin_config_schematic}). It is interesting to follow the fate of DWs under the DSF dynamics. Table \ref{table:dsf_moves} shows that (a) zero-energy moves correspond to a single DW moving by a distance $2a$ in each step, where $a$ denotes the lattice spacing, (b) if two DWs are a distance $2a$ apart, a DSF move which flips the intervening pair of spins leads to the annihilation of both DWs and a concomitant lowering of the energy, (c) if two DWs are a distance 3$a$ apart, they cannot approach any closer. 

Now consider two successive DWs as shown schematically in Fig. \ref{spin_config_schematic}. Two cases arise. (i) If the separation between DWs is an even multiple of $a$, then they get annihilated at a long time. (ii) If the DWs are separated by an odd multiple of $a$, then it is not possible to annihilate these DWs, as shown in more detail in Sec. \ref{coarsening}.

In view of (i) and (ii) above, we deduce that a succession of annihilations leads to the steady state with $N$ domains, each containing an odd number of parallel spins; further, the minimum size of each domain is three.  The question then arises: What is the typical value of $N$ in the steady state of a system of size $L$? We observe that $N$ is equal to the sublattice magnetization $\mathcal{M}\equiv M_{\mathrm{even}}-M_{\mathrm{odd}}$, defined in Section \ref{modeldetails}, as each domain contributes unity to $\mathcal{M}$. Since $\mathcal{M}$ is conserved by DSF dynamics its value can be deduced from the initial configuration. Since the initial state is purely random, $\mathcal{M}$ has a binomial distribution from $-L$ to $L$ with zero mean and standard deviation $\sqrt{L}$ for a system of size $L$. Therefore, the typical number of domains $N$ in steady state is $\sim\mathcal{O}(\sqrt{L})$. 
On the other hand, if we start with a carefully designed random initial state with $\M=0$, the system reaches a state with all spins either up or down  and the dynamical rules in this regime at $T=0$ forbid any dynamics in this state. We show the evolution of such a state in appendix \ref{slmzero}.

Thus we arrive at the following simple description of the steady state. A conserved number $N$ of DWs separate $N$ domains, each with an odd number $ \ge 3$ of parallel spins. Each DW performs a random walk with a step length of $2a$ with the ‘hard core’ constraint that the closest distance of approach of two successive walkers is $3a$. The total number $\Omega(N,L)$ of allowed configurations with $N$ walkers in a system of size $L$ can be found by considering the distribution of $L$ spins in $N$ boxes, ensuring an odd number, at least 3, in each box. The result is $\Omega(N,L) =(2L/N){(L-N-2)/2 \choose (N-1)}$.

We now invoke the condition of detailed balance to show that in steady state, every allowed configuration $\mC$ is equally likely, and is given by $ P_{\text{ss}}(\mC) = 1/\Omega(N,L) $.
Since all allowed zero-energy moves occur at the same rate (Table \ref{table:dsf_moves}), we have $W(\mC\rightarrow \mC^{\prime})=W(\mC^{\prime}\rightarrow \mC)$, where $W(\mC\to \mC')$ is the rate from configuration $\mC$ to $\mC'$. This implies $P_{\text{ss}}(\mC)W(\mC\rightarrow  \mC^{\prime}) = P_{\text{ss}}(\mC^{\prime})W(\mC^{\prime}\rightarrow \mC)$, which proves the assertion above.

\begin{figure}
	\includegraphics[width=8.6cm]{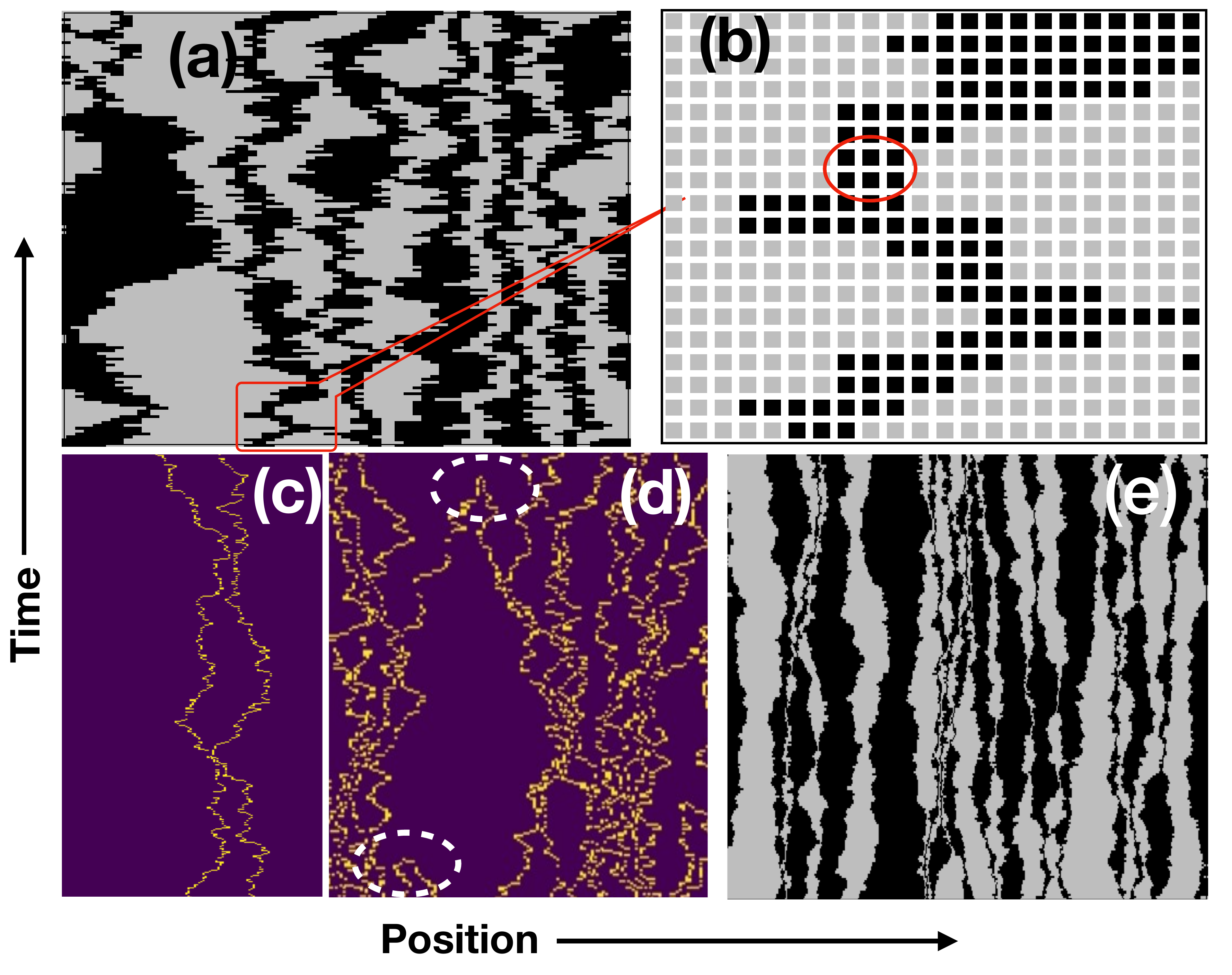}
	\caption{In all the subfigures, the vertical axis denotes increasing time and the horizontal axis denotes increasing site index. (a) A typical timeline of the arrested state, where black pixels are down-spins, gray pixels are up-spins. (b) A zoomed-in version of the timeline. (c) A timeline showing the hard-core interaction between walkers, where yellow pixels are the domain walls, purple pixels are domains. (d) The system is still coarsening. Some annihilations of the domain walls are circled, where yellow pixels are the domain walls, purple pixels are domains. (e) A typical timeline of the SEP-mappable Ising system, where black pixels are down-spins, gray pixels are up-spins.} 
	\label{figure:schematic}
\end{figure}

Figure \ref{figure:schematic} illustrates features of the steady state and the approach to it. Figure \ref{figure:schematic}(a) shows the time evolution of the system in steady state on a large scale. The zoomed-in version in Fig. \ref{figure:schematic}(b) and successive collisions of two walkers (domain walls) shown in Fig. \ref{figure:schematic}(c) illustrate the $3a$- hard core constraint which operates between successive walkers.  Figure \ref{figure:schematic}(d) shows the evolution of the system during the approach to steady state, when annihilations of successive even-separation DWs occur. Finally, Fig. \ref{figure:schematic}(e) shows the time evolution when DWs follow the rules of the Simple Exclusion Process (SEP) \cite{spitzer1970,liggettbook}. We will argue below that our system resembles the SEP, which can then be used to deduce the time dependence of correlation functions at long times.

\section{Results}
\label{results}

We now present the results of numerical simulations which quantitatively characterize the state with mobile domain walls. We begin with one of two initial configurations: 

($\mC_1$) A configuration with spins at every site chosen at random, but subject to the constraint that the sublattice magnetization is $\mathcal{M}$, with $\mathcal{M}$ even. The number of DWs in such a configuration is typically much larger than $\mathcal{M}$.

($\mC_2$) A configuration with an even number $\mathcal{M}$ of domain walls placed at random, but subject to the constraint that every domain has an odd number $n$ of spins with $n\geq 3$. As shown in Section \ref{domainwalls}, every configuration of this type occurs with equal probability in steady state.

Configurations of type $\mC_1$ are used for studies of coarsening, while configurations of type $\mC_2$ are appropriate for studies of dynamics within steady state. At large enough times, an evolution starting with a $\mC_1$ configuration leads to a steady state configuration of type $\mC_2$, as verified also by numerical simulations. In the simulations discussed below, we chose $\mathcal{M}=\sqrt{L}$, typical of a purely random state.

 \subsection{Static Correlation Function in Steady State}
\begin{figure}
\includegraphics[width=8.6cm]{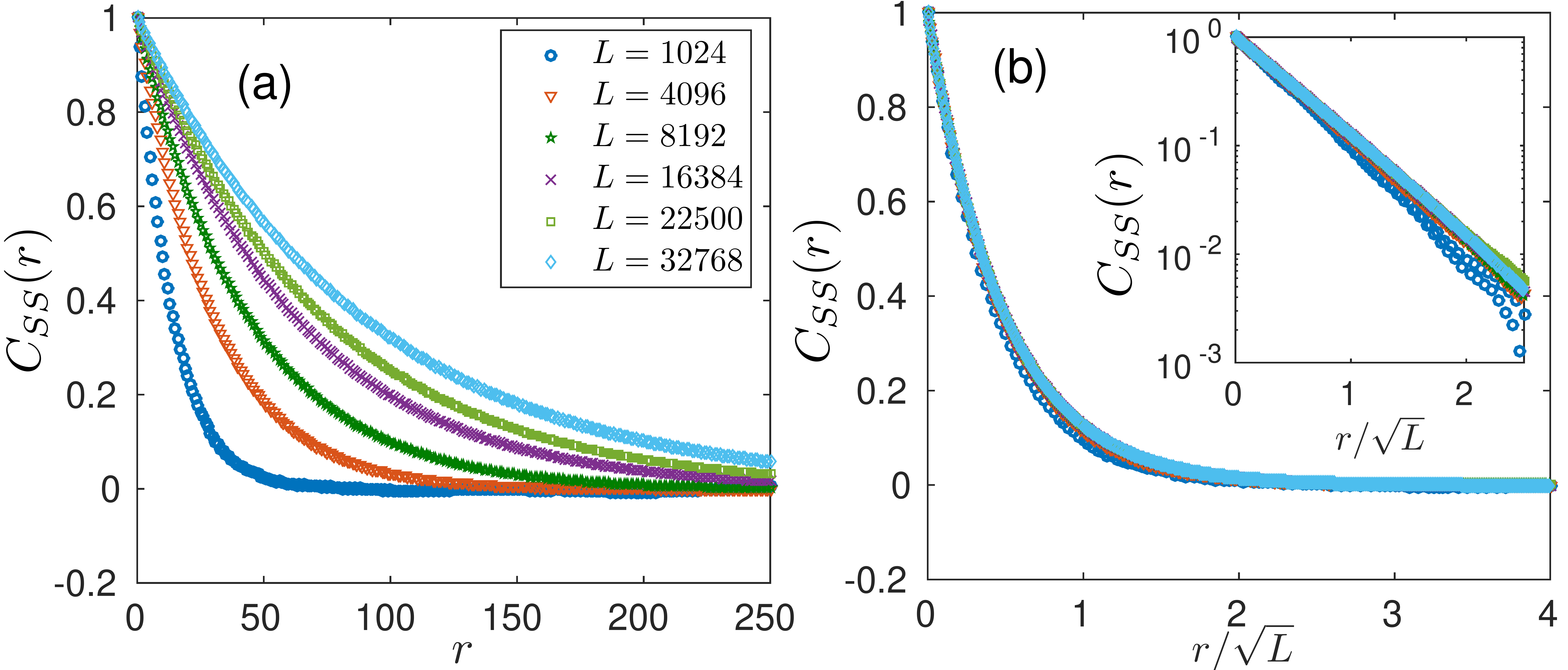}
\caption{Spin-spin spatial correlation functions in the steady-state follow $\exp(-2r/\sqrt{L})$. (a) $C_{SS}(r)$ vs $r$ in linear scale. (b) Scaling collapse. $C_{SS}(r)$ vs $r/\sqrt{L}$. Inset in (b): $C_{SS}(r)$ vs $r/\sqrt{L}$ in semi-log scale. Each point is an average over 5000 histories.}
\label{figure:ss_spatial_corr} 
\end{figure}

As discussed above, we begin with a configuration of type $\mC_2$ with $\mathcal{M}=\sqrt{L}$. The correlation function in the steady-state is defined as
\begin{equation}
 C_{SS}(r)=\frac{1}{L}\sum_{i=1}^L\left\langle S_{i}S_{i+r}\right\rangle,
\end{equation}
where $\langle\ldots\rangle$ denotes a steady-state ensemble average. $C_{SS}(r)$ is plotted in Fig. \ref{figure:ss_spatial_corr}(a) for different system sizes $L$. Due to the presence of the $\sqrt{L}$-sized domains in the steady-state, we expect $C_{SS}(r)$ to decay exponentially with a length-scale $\sqrt{L}$. Indeed, as shown in Fig. \ref{figure:ss_spatial_corr}(b), plotting $C_{SS}(r)$ as a function of $r/\sqrt{L}$ leads to an excellent scaling collapse to a single master curve. The inset in Fig. \ref{figure:ss_spatial_corr}(b) shows a straight line for the same plot on a semi-log scale; this confirms the exponential behavior. As we show in Sec. \ref{mappingtoSEP}, by relating the large-distance, long-time properties in the steady-state of our model to that of a simple exclusion process (SEP), we find
\begin{equation}
C_{SS}(r)=\exp\left(-\frac{2r}{\sqrt{L}}\right).
\label{eqn:c_r_ss}
\end{equation}
Our simulation data supports this analytical prediction. 

Note in passing that our system displays long-range order (LRO) defined by 
\begin{equation}\label{lro}
m_{0}^{2}\equiv\lim_{r\rightarrow\infty}\lim_{L\rightarrow\infty}\left\langle S_{i}S_{i+r}\right\rangle.
\end{equation}
Although ordered regions are not of order system size $L$ as is customary, the fact that they are of order $\sqrt{L}$ suffices to ensure that $m_0^2=1$.

\subsection{Correlation function in coarsening regime}
\label{coarsening}
We next look at the coarsening following a sudden quench to $T=0$ starting at $t=0$ with random initial configuration of type $\mC_1$. Domain walls separated by ordered regions with an even number of spins diffuse until they are two lattice spacings apart, at which point they annihilate. Finally, the system approaches the steady state [see Fig. \ref{figure:schematic}(d)] with $\mathcal{M}\sim \sqrt{L}$ domain walls. 

\begin{figure}
	\includegraphics[width=8.6cm]{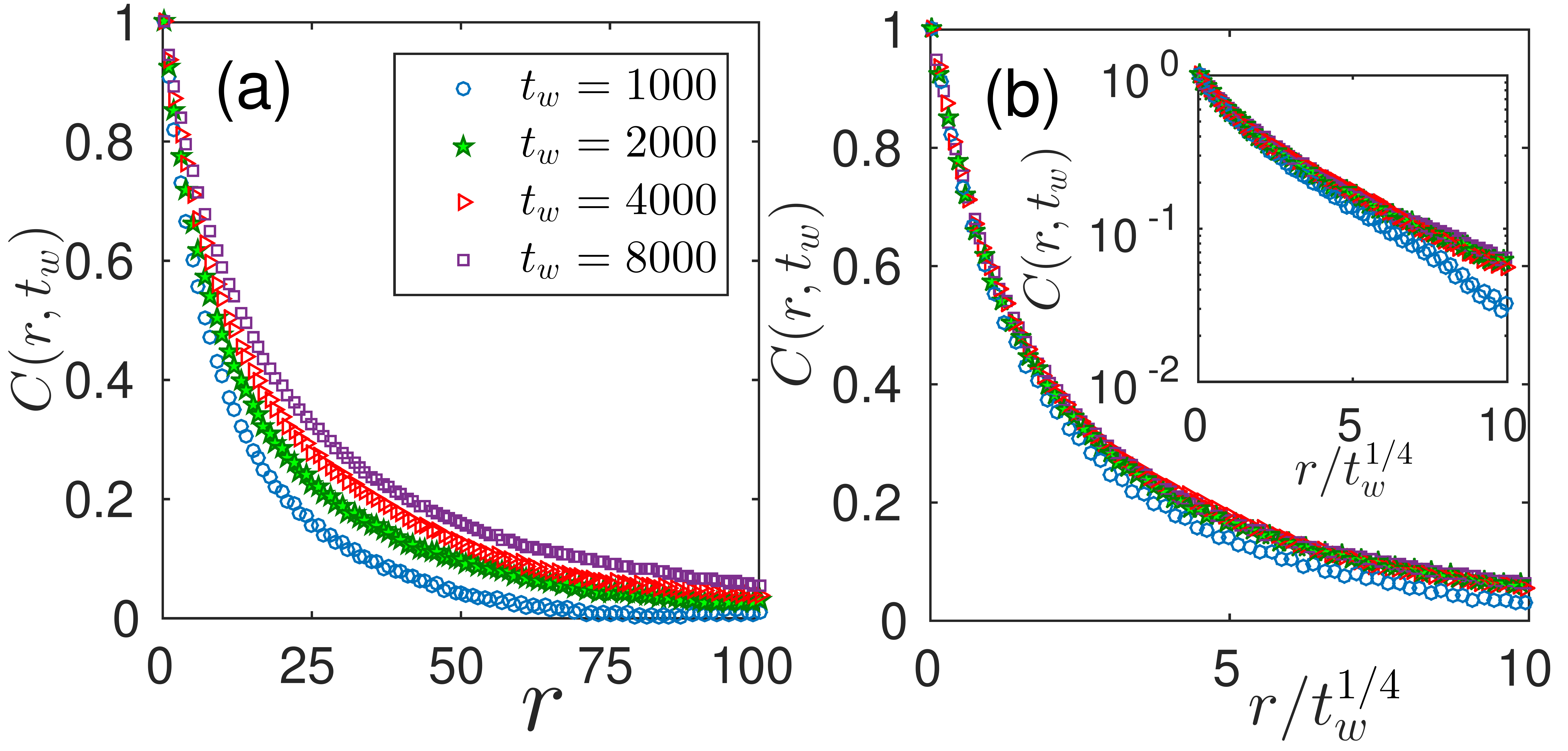}
	\caption{Spin-spin spatial correlation functions in the coarsening regime follow $\exp(-\Gamma r/t_w^{1/4})$. (a) $C(r,t_w)$ vs $r$ in linear scale. (b) Scaling collapse. $C(r,t_w)$ vs $r/{{t_w}^{1/4}}$. Inset in (b): $C(r,t_w)$ vs $r/{{t_w}^{1/4}}$ in semi-log scale. Each point is an average over 1000 histories in a system with L=4096.}
	\label{coarseningplot}
\end{figure}

In the coarsening regime, there are two length scales of relevance, each growing with a different power of the waiting time $t_w$. The first is $\mathcal{L}(t_w)$, the length scale below which the system has achieved a quasi-equilibrium. Here $\mathcal{L}(t_w)$ is the typical separation between two even-sized domains, and considering the diffusive nature of the process which leads to the annihilation of even domains, we have $\mathcal{L}(t_w) \sim t_w ^ {1/2}$. Now each such equilibrated stretch holds a large number of DWs separating odd-sized domains. The second length scale of relevance is $\mathcal{R}(t_w)$, the mean separation of these DWs, which scales as $\sqrt{\mathcal{L}(t_w)}$  and determines the decay of the two-point correlation function. Evidently we have $\mathcal{R}(t_w) \sim t_w^ {1/4}$, so we expect the correlation function in the coarsening regime to follow
\begin{equation}\label{coarsening_scaling}
C(r,t_w)=\exp\left(-\frac{\Gamma r}{{t_w}^{1/4}}\right),
\end{equation}
where $\Gamma$ is a constant.
Figure \ref{coarseningplot}(a) shows the correlation function $C(r,t_w)$ as a function of $r$ after a quench from the random configuration to $T=0$ for four different values of $t_w$ for a system with $L=4096$. Figure \ref{coarseningplot}(b) shows that when $C(r,t_w)$ is plotted as a function of $r/t_w^{1/4}$, a very good data collapse is obtained at large values of $t_w$ confirming the scaling prediction, Eq. (\ref{coarsening_scaling}). The inset of Fig. \ref{coarseningplot}(b) shows the same plot in semi-log scale, clearly showing a better collapse for larger values of $t_w$. We provide a detailed description of the coarsening dynamics in Appendix \ref{coarsening_details}.

\subsection{Auto-Correlation Function in Steady-State}
\label{autocorr_sec}
The spin-spin auto-correlation function, $\phi(t)$, in steady-state is defined as
\begin{equation}
\phi(t)\equiv\frac{1}{L}\sum_{i}\left\langle S_{i}(t_0)S_{i}(t_0+t)\right\rangle -\left\langle S_{i}(t_0)\right\rangle ^{2},
\end{equation}
where $\langle\ldots\rangle$ represents an average over $t_0$. 
As already discussed,  in steady state, there are $\mathcal{M}=(\sqrt{L})$ domains and an equal number of DWs, with a hard core repulsion between successive DWs [see Fig. \ref{figure:schematic}(c)]. Each DW performs a random walk through the energy-conserving moves alone as listed in Table \ref{table:dsf_moves}. This feature leads to the conservation of $\mathcal{M}$, a dynamical constraint in the steady-state.

\begin{figure}
	\includegraphics[width=8.6cm]{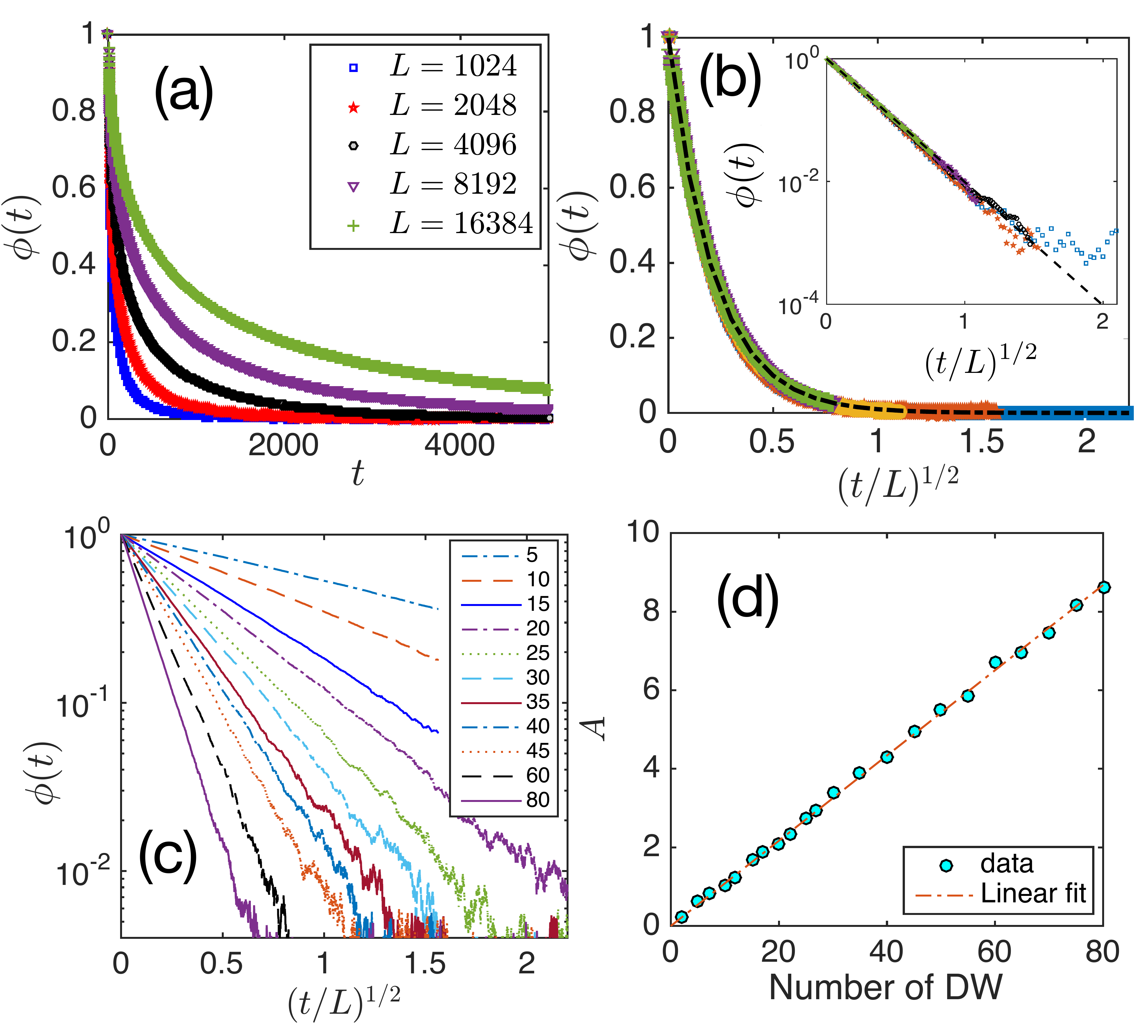}
	\caption{Spin-spin auto-correlation functions in steady-state follow $\exp(-A (t/L)^{1/2})$. (a) $\phi(t)$ vs $t$ in linear scale in the steady state with exactly $\sqrt{L}$ domain walls. (b) Scaling collapse. $\phi(t)$ vs $(t/L)^{1/2}$ for the same set of data presented in (a). The line is a fit of Eq. (\ref{eqn:c_t_ss}) that yields $A=4.6$. {\bf Inset:} $\phi(t)$ vs ${(t/L)}^\frac{1}{2}$ in semi-log scale. Each point is an average over at least 5000 histories and in some cases it is substantially larger, for instance $10^5$ for the smallest $L$. (c) Evolution of $\phi(t)$ as function of $(t/L)^{1/2}$ with different number of DWs as quoted in the legend and $L=2048$. (d) The coefficient $A$ in Eq. (\ref{eqn:c_t_ss}) increases linearly with the number of DWs. Symbols are results obtained from fitting Eq. (\ref{eqn:c_t_ss}) with simulation data of a system of size $L=2048$ and the line is a fit $A(x)=\kappa x$ with $\kappa=0.11$.}
	\label{figure:ss_auto_corr}
\end{figure}

 From simulation, we find that $\phi(t)$ shows stretched exponential relaxation in steady-state,
\begin{equation}
\phi(t)=\exp\left(-A\left(\frac{t}{L}\right)^{\frac{1}{2}}\right)
\label{eqn:c_t_ss}
\end{equation}
where $A$ is a constant. We plot $\phi(t)$ as function of $t$ for different system sizes $L$ in Fig. \ref{figure:ss_auto_corr}(a). We obtain excellent data collapse when we plot $\phi(t)$ as function of $(t/L)^{1/2}$ as shown in Fig. \ref{figure:ss_auto_corr}(b). The inset shows the same data on a semi-log scale where it is expected to be a straight line at large $(t/L)^{1/2}$. 
The coeffcient $A$ in Eq. (\ref{eqn:c_t_ss}) depends on the number $N$ of DWs present in the system; our simulations show that $A$ varies linearly with $N$ (Fig. \ref{figure:ss_auto_corr}d). Since the DSF dynamics does not annihilate DWs in the steady state, $N$ is determined by the sublattice magnetization $\M$ of the random initial state. For convenience of simulation, we generate steady state configurations with precisely $N=\sqrt{L}$ DWs and take an average over these configurations, which is equivalent to averaging over $t_0$.
From a fit of Eq. (\ref{eqn:c_t_ss}) with the simulation data we obtain $A\simeq 4.6$.

\begin{figure}
\includegraphics[width=8.6cm]{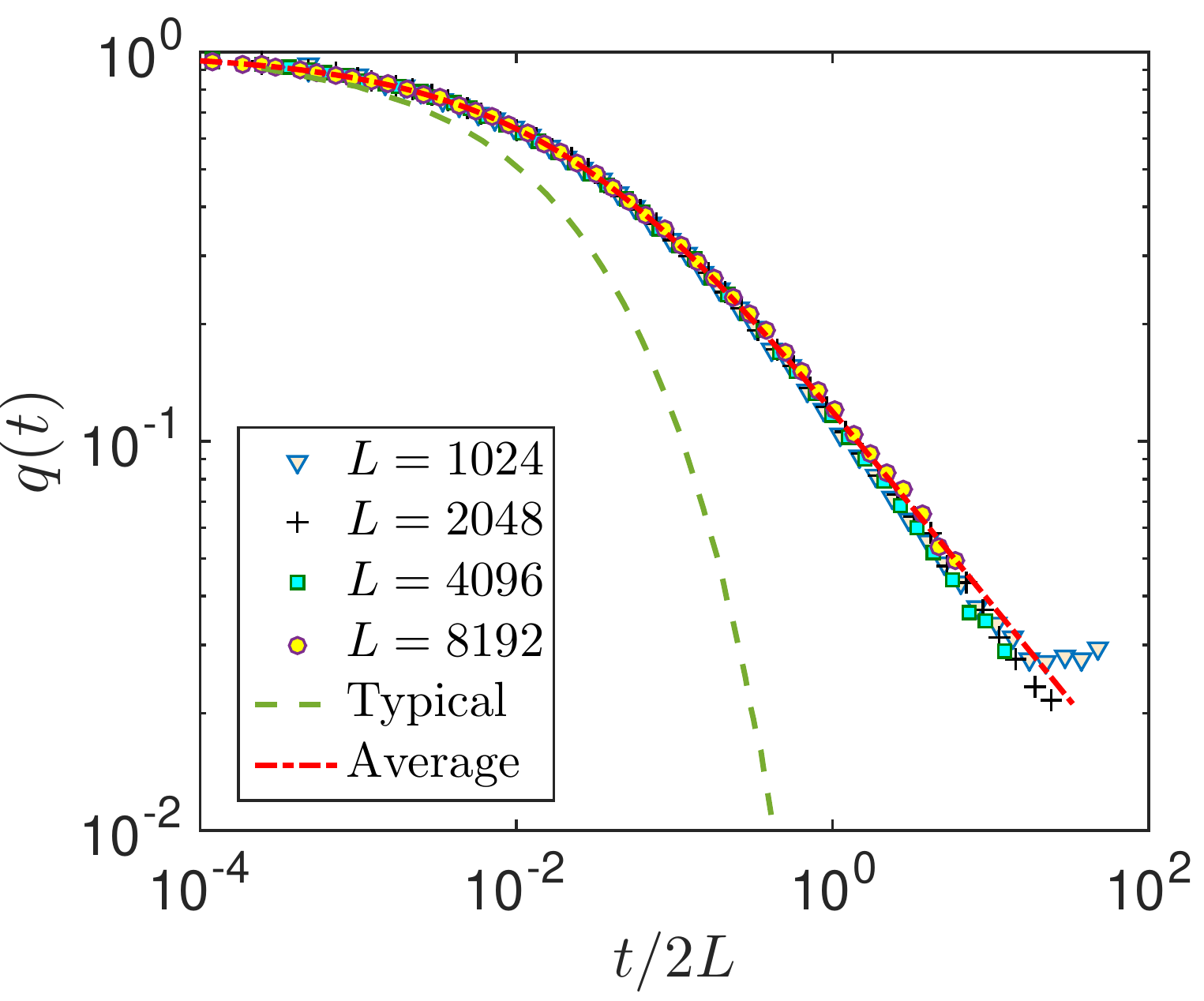}
\caption{Simulation data for the autocorrelation function $q(t)$ averaged over the ensemble of initial states, plotted for different $L$ as noted in the figure. Equation (\ref{ensavqoft}) gives the form of a master curve when $q(t)$ is plotted as a function of $t/2L$. The figure shows that the numerical data agrees well with the plot of Eq. (\ref{ensavqoft}) with $A_0 = 4.6$. Also plotted in the figure is the autocorrelation function $\phi(t)$ starting with a single, typical configuration with $L=2048$ and $\sqrt{L}$  DWs. Notice that $q(t)$ decays much slower than $\phi(t)$ at long times. Each point is an average over at least 5000 histories}.
\label{qoftplot}
\end{figure}

\subsection{Ensemble average of autocorrelation function}
Consider performing successive quenches from the $T= \infty$ state, and averaging the auto-correlation function over the ensemble of steady states reached, which in general have different numbers $N$ of DWs. Since the number of DWs in steady state is conserved, it follows that an average over this ensemble and an average over $t_0$  for a particular value of $N$ need not be the same. In fact, as we show below, there are significant differences, especially at large times.
	
We define the ensemble average of the autocorrelation function, $q(t)$, in steady state,
	\begin{equation}
	q(t)=\overline{\phi(t)}=\overline{\frac{1}{L}\sum_{i}\left\langle S_{i}(t_0)S_{i}(t_0+t)\right\rangle -\left\langle S_{i}(t_0)\right\rangle ^{2}},
	\end{equation}
where $\overline{\mathcal{O}}$ defines ensemble average of $\mathcal{O}$. Starting from a random initial condition, the number of DWs in steady state has a distribution, given as
\begin{equation}\label{dwdist}
P(N)=\sqrt{\f{2}{\pi L}}\exp{\left[-\f{N^2}{2L}\right]}
\end{equation}
where $N$ ($\geq0$) is the number of DWs. We saw in Sec. (\ref{autocorr_sec}) that the coefficient $A$ in Eq. (\ref{eqn:c_t_ss}) increases linearly with the number of DWs, whose typical number is of the order of $\sqrt{L}$ as can be seen from Eq. (\ref{dwdist}). Therefore, we write $A=A_0 N/\sqrt{L}$ where $A_0$ is the typical value of $A$. Averaging over $N$, we obtain $q(t)$ as
\begin{align}\label{ensavqoft}
q(t)&=\sqrt{\f{2}{\pi L}}\int_0^\infty\exp{\left[-\f{N^2}{2L}\right]}\exp{\left[-\f{A_0Nt^{1/2}}{L}\right]} \nonumber\\
&=\exp\left[\f{A_0^2t}{2L}\right]\text{erfc}\left(A_0\sqrt{\f{t}{2L}}\right).
\end{align}
At short times $t<<L$ , Eq. (\ref{ensavqoft}) reduces to $q(t) \approx (1- A_0\sqrt{2/\pi}(t/L)^{1/2})$, which is consistent with Eq. (\ref{eqn:c_t_ss}). On the other hand, at long times $t>>L$ , we may use the asymptotic form $\text{erfc}(x)\sim \exp{(-x^2)}/\sqrt{\pi}x$ to obtain $q(t) \approx \sqrt{2/\pi A_0^2}(t/L)^{-1/2}$. Figure \ref{qoftplot} shows very good agreement of Eq. (\ref{ensavqoft}) with simulation results for $q(t)$. For comparison, the figure also shows the decay of $\phi(t)$ for a single typical configuration with $L = 2048$ and $\sqrt{L}$ number of DWs. We see that there is a significant difference at long times: while $\phi(t)$ exhibits SSER throughout, $q(t)$ shows a slow power-law decay at long times $t$. This difference arises as at large $t$, there is always a fraction of members of the ensemble (those with $A_0N/L^2<< t$) which would have decayed very little, and would contribute strongly to $q(t)$.

\section{Relation to the Simple Exclusion Process}
\label{mappingtoSEP}
From the coarse-grained point of view, in the steady state the system consists of $\mathcal{M}=\sqrt{L}$ domains separated by diffusing DWs with a no-crossing constraint. We expect that microscopic details, like the odd number of spins in each domain, the domain walls moving in steps of two lattice spacings and the minimum possible distance between the walls being three lattice spacings should not matter for large-separation or long-time properties. 

The DW dynamics should then be well approximated by the simple exclusion process (SEP) that describes a system of identical particles with hard-core repulsion moving on a lattice \cite{spitzer1970,liggettbook}. Each particle attempts to move to its nearest neighbor sites with equal probability; the move is implemented if the site in question is not already occupied. 
The dynamics in our arrested system in steady-state resembles that of an Ising spin chain with a number $\sqrt{L}$ of alternating domains, evolving with the rule that a single spin can flip only if the two neighboring spins are mutually antiparallel. 
Such spins are necessarily at the boundaries of the domains. The spin system is related to the SEP through the relation $\eta_{i+1/2}(t)=\frac{1}{4}(S_i(t)-S_{i+1}(t))^2$ where $\eta_{i+1/2}=1$ if there is a DW (which represents a particle in the SEP) between sites $i$ and $i+1$, and $0$ otherwise. 
A particular snapshot of the simulation of this system is presented in Fig. \ref{figure:schematic}(e). As expected, with increasing system size, the details of the original arrested steady-state become irrelevant and the behaviors of the original and the new system converge. Thus, we consider a system that has $\sqrt{L}$ DWs which perform random walks but cannot cross each other -- a SEP with a subextensive number of particles. 


A similar system was considered in Refs. \cite{skinner1983}, \cite{spohn1989} and \cite{godreche2015}, but with the important difference that the number of DWs was proportional to $L$. In that case, the autocorrelation function follows $\phi(t)\sim \exp(-(t/\tau)^{1/2})$ where upper and lower bounds derived in \cite{spohn1989} imply that $\tau$ remains finite in the limit $L\to\infty$. For the system of interest in this work, the density of domain walls, $\rho=1/\sqrt{L}\to0$  as $L\rightarrow\infty$, which leads to $\tau$ being proportional $L$, while the exponent $\beta$ of the SER as well as the linear-dependence of $A$ in Eq. (\ref{eqn:c_t_ss}) on the number of DWs remain the same. 

We now turn to the analytic derivations of the spatial correlation and temporal auto-correlation functions in the steady-state, using the independent interval approximation (IIA) \cite{majumdar1996,barma2019}.

{\it Spatial correlation function:} The spatial correlation function, $C_{SS}(r)$ can be written as
\begin{equation}\label{cssr_def}
C_{SS}(r)=\sum_{n=0}^{\infty}(-1)^n p_n(r),
\end{equation}
where $p_n(r)$ is the probability that a segment of length $r$ has exactly $n$ domain walls. If domains are drawn independently from a common distribution $P(l)$, the correlation function in Laplace space, $\tilde{C}(s)=\int^{\infty}_0{C_{SS}(r)e^{-sr}\mathrm{d}r}$, can be written as 
\begin{equation}
\tilde{C}(s)=\frac{1}{s}-\frac{2}{\langle{l}\rangle{s^2}}\frac{1-\tilde{P}(s)}{1+\tilde{P}(s)},
\label{eqn:iia_relation}
\end{equation}
where $\tilde{P}(s)$ is the Laplace transform of $P(l)$ and $\langle{l}\rangle$ is the average domain size.
For large $L$, the probability that a domain is of length $l$ is given by $P(l)={(1-\rho)^l}\rho$ where $\rho=\sqrt{L}/L=1/\sqrt{L}$ is the density of the domain walls. For large $l$, $P(l) \approx {\rho e^{-\rho l}}$, which gives us $\langle{l}\rangle = \sqrt{L}$ and $\tilde{P}(s)=\frac{\rho}{s+\rho}$. Using these relations in Eq. \ref{eqn:iia_relation}, we get
\begin{equation}
\tilde{C}(s)=\frac{1}{s+\frac{2}{\sqrt{L}}}.
\end{equation}
Inverting the Laplace transform gives us \eqref{eqn:c_r_ss}.

{\it Auto-correlation function:} 
We now argue that the temporal autocorrelation function $\phi(t)$ can be found similarly by focusing on the time segments between successive returns of a DW, noting that each sign change of the spin is associated with such a return of the DW. Thus in analogy with Eq. (\ref{cssr_def}), we write
\begin{equation}
\phi(t) = \sum_{n=0}^\infty (-1)^n p_n(t)
\end{equation}
where $p_n(t)$ now stands for the probability that there are $n$ returns to the origin in a time stretch $t$. The probability that a random-walking DW first returns to its starting point at time $t$ follows $P_{return}(t)  \sim  t^{-3/2}$ \cite{feller1968}, assuming that the DW performs an unhindered random walk. This would be expected to hold up to a time $t_{coll}$, which is of order the typical collision time between successive walkers. Since the mean spacing between walkers is $\sqrt{L}$, we expect $t_{coll}$ to be proportional to $L$.

Since successive returns to the origin for a free random walk are independent events, within the time $t_{coll} \sim L$, we use the IIA to make further progress. Note that the mean return time $\tau$ within $t_{coll}$ is $\tau=2\sqrt{L}$ and that the Laplace transform of $P_{return}(t)$ is $\tilde{P}(s) \sim 1 +\Gamma(–\f{1}{2}) \sqrt {s}$. We now use Eq. (\ref{eqn:iia_relation}),  replacing $\tilde{C}(s)$ by $\tilde{\phi}(s)$, (the Laplace transform of $\phi(t)$)  and $\langle l\rangle$ by $\tau$.  We obtain the small-$s$ behaviour  of $\tilde{\phi}(s)$ as
\begin{equation}
\tilde{\phi}(s)\approx \frac{1}{s}- \frac{2\sqrt{\pi}}{\tau}s^{-3/2}.
\end{equation}
Performing the inverse Laplace transform, we find
\begin{equation}
\phi (t) \approx 1 - 2\left(\frac{t}{L}\right)^{1/2}
\end{equation}
which, for small values of $t/L$, coincides with the size-stretched exponential form Eq. (\ref{eqn:c_t_ss}).

\section{Conclusion}
\label{disc}

We have studied the relaxation properties of a system in an arrested state in which the dynamics involves a conserved number of interacting, diffusing excitations. The key point which underlies the unusual behavior of the system is that the number of excitations grows sub-extensively with the system size. This ultimately leads to the feature of SSER in the autocorrelation function in steady state, namely stretched exponential decay with a relaxation time that diverges with growing system size. It also leads to an unusual feature in the coarsening dynamics describing the approach to the steady state, namely the existence of two length scales, each growing with a different power of the time. Further, the characteristic time-dependent behavior of a single typical sample differs markedly from the average behaviour over an ensemble of initial conditions.

We demonstrated these properties in the ANNNI model with double spin flip dynamics that conserves the staggered magnetisation. For a range of parameters, a rapid quench from a disordered state towards zero temperature leads to an arrested steady state with a conserved number of domain walls separating domains with an odd number of parallel spins. In the steady state, the dynamics is governed by energy-conserving moves, under which domain walls diffuse but respect a no-crossing constraint. In a system of size $L$, we showed that the spatial correlation function decays exponentially with a length scale that varies as $\sqrt{L}$ and the system exhibits long range order. The auto-correlation function in the steady state displays a stretched-exponential relaxation with the stretching exponent $\beta = 1/2$. The noteworthy feature is that the relaxation time depends on $L$. The form of relaxation of a single typical sample in steady state differs markedly from that averaged over an ensemble of initial conditions resulting from different quenches. The latter shows a slow power law decay at large times, reflecting the propensity of systems with a low value of $N$ to relax very slowly.

In the coarsening regime, the size of odd-spin domains grows as $t_w^{1/4}$, whereas the typical separation between even-sized domains grows as $t_w^{1/2}$. This pattern of growth continues indefinitely in an infinite system, but in a finite system of size $L$, as $t_w \to \infty$, the number of even sized domains vanishes, while the number of odd-sized domains approaches $\sqrt{L}$, characteristic of the steady state.

For large-distance long-time properties in the steady state, our system resembles an Ising system where the dynamics involves only energy conserving moves \cite{skinner1983,spohn1989,godreche2015}, insofar as domain walls move diffusively in the steady state and respect a no-crossing constraint in both systems. But the analogy cannot be extended too far. For instance, in the coarsening regime, the occurrence of two distinct diverging length scales in our system (in contrast to the single diverging length in the Ising model) can be traced ultimately to the competing interactions of the ANNNI model along with the conservation law implied by double spin-flip dynamics.

The dynamics in the arrested steady state studied in this paper is constrained since we consider dynamic evolution at $T = 0$ and allow only energy-conserving moves. In this sense, our system belongs to the class of kinetically constrained models (KCMs) that have been studied earlier in the context of glass. Like several other KCMs, our system too shows stretched-exponential relaxation. But there are a couple of distinctive features:  first, the relaxation time diverges as a power of the system size, corresponding to SSER; and secondly, the average over the ensemble of initial conditions yields a form of relaxation which is quite different from that of a single sample. 
It would be interesting to identify such features in other KCMs as well.


\section{Acknowledgements}
This project was funded by intramural funds at TIFR Hyderabad from the Department of Atomic Energy (DAE), Government of India. VG thanks TIFR Hyderabad for hospitality and support. MB acknowledges support under the DAE Homi Bhabha Chair Professorship of the Department of Atomic Energy.

\begin{figure*}
	\includegraphics[width=14.6cm]{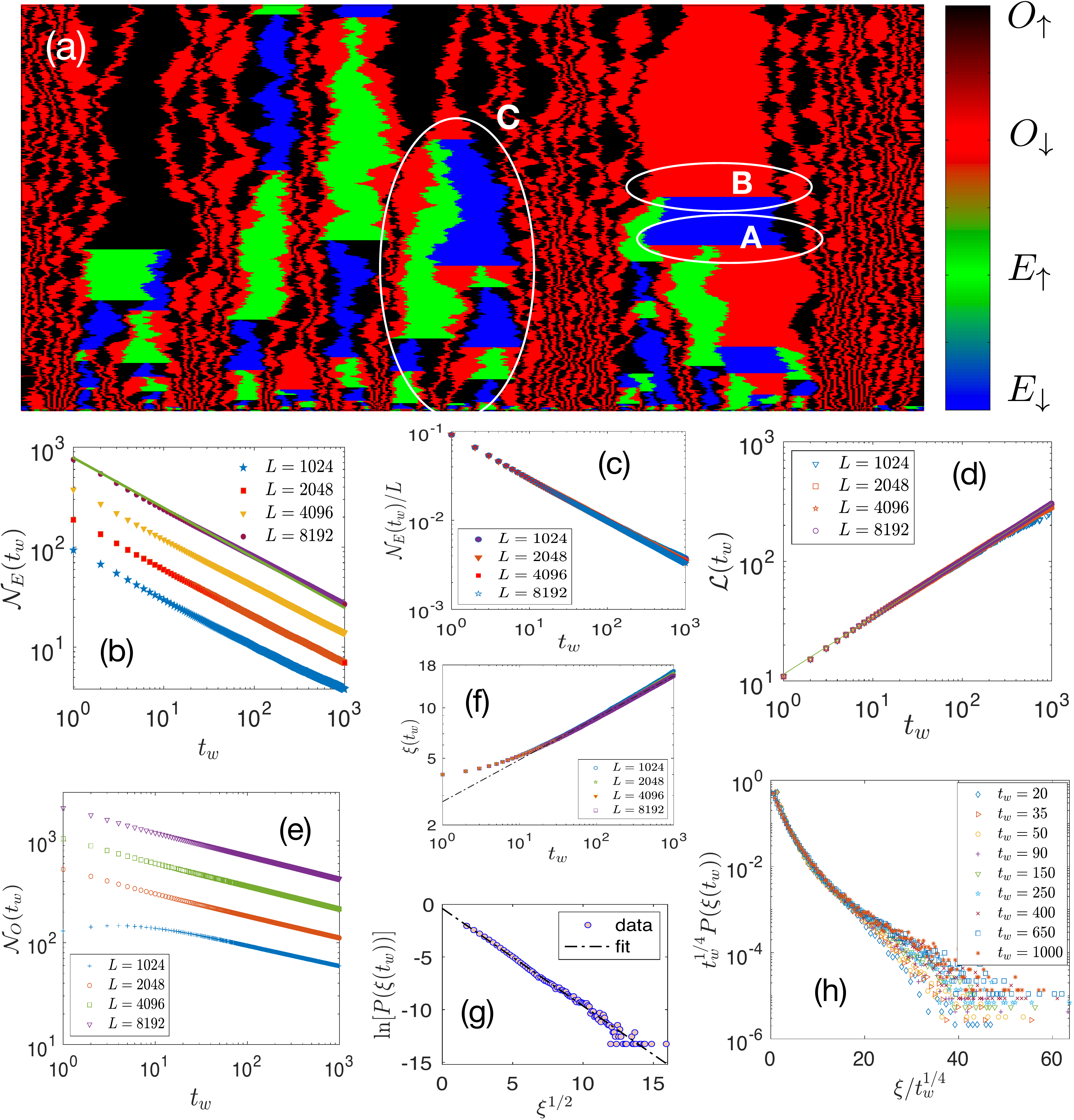}
	\caption{Details of coarsening, following even and odd spin clusters. (a) Time evolution (upward) towards the steady state, starting from a random configuration. Black and red colors mark odd domains with spins $\uparrow$ and $\downarrow$, denoted as $\Ou$ and $\Od$ respectively. Green and blue colors show even domains with $\uparrow$ and $\downarrow$, denoted as $\Eu$ and $\Ed$ respectively. The highlighted regions, A, B and C show three mechanisms discussed in points (iv), (v) and (vi) in the main text. (b) The number of even domains $\mathcal{N}_E(t_w)$, decreases as $t_w^{-1/2}$, the line being a fit with a function $f(x) \sim x^{-1/2}$. (c) During coarsening in a system of length $L$, $\mathcal{N}_E(t_w)$ is expected to be proportional to $L$. $\mathcal{N}_E(t_w)/L$ for different $L$ follows a master curve. (d) $\mathcal{L}(t_w)$ grows as $t_w^{1/2}$, the line being a fit. (e) Number of odd domains, $\mathcal{N}_O(t_w)$ decreases as $t_w^{-1/4}$. (f) Length of odd domains increases as $t_w^{1/4}$. (g) The fit shows that $P(\xi)\sim \exp(-\xi^{1/2})$ and thus, plot of $\ln[P(\xi)]$ as a function of $\xi^{1/2}$ becomes linear. (h) Scaling collapse of the distribution of odd domain sizes confirm that odd domain length increases as $t_w^{1/4}$. }
	\label{coarsening_characteristics}
\end{figure*}

\appendix
\section{Details of the coarsening dynamics}
\label{coarsening_details}
Below we discuss key characteristics of the approach to the steady-state, based on a detailed consideration of domain dynamics.
As discussed in Sec. \ref{domainwalls}, domains with an even number of spins (denoted $E$) get annihilated at long times, and in steady state we are left only with domains with an odd number of spins (denoted $O$). Even domains with $\uparrow$ and $\downarrow$ spins are denoted as $E_\uparrow$ and $E_{\downarrow}$ and odd domains with $\uparrow$ and $\downarrow$ spins as $O_\uparrow$ and $O_{\downarrow}$ respectively. The time evolution starting from an arbitrary initial configuration is shown in Fig. \ref{coarsening_characteristics}a, where black and red refer to $\Ou$ and $\Od$ respectively, and green and blue refer to $\Eu$ and $\Ed$ respectively.

We now turn to a discussion of domain dynamics.

\begin{figure}
	\includegraphics[width=8.6cm]{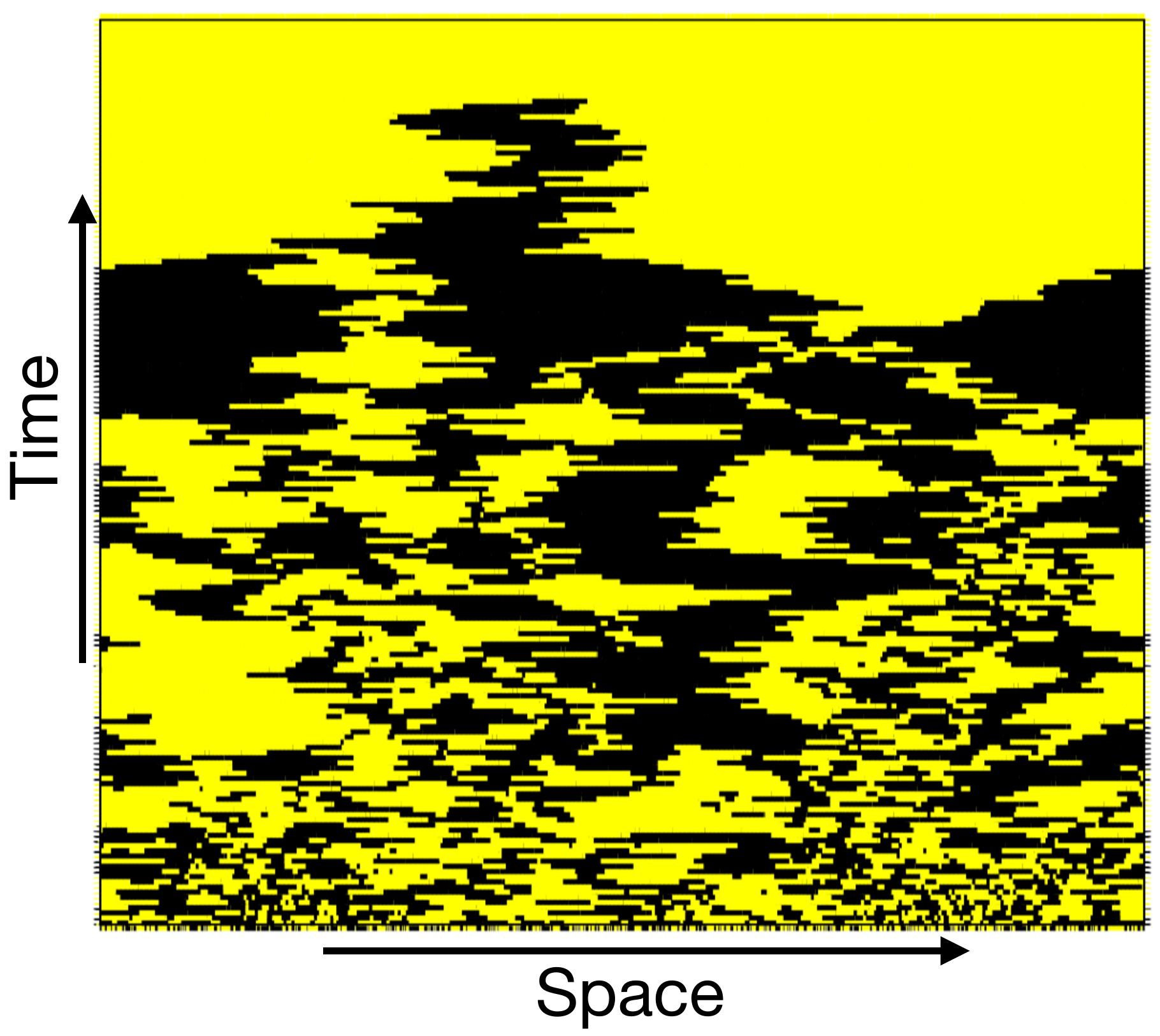}
	\caption{Time evolution of the configuration for a quench to $T=0$ starting from a random initial state with $M_{even}=M_{odd}$. The color black refers to $\downarrow$ and yellow to $\uparrow$ with system size $L=1024$ and $M_{even}=32$.}
	\label{slmzeroevolution}
\end{figure}

(i) The number of $E$ domains, $\mathcal{N}_{E}(t_w)$, decreases as $t_w^{-1/2}$ as shown in Fig. \ref{coarsening_characteristics}(b) for four different system sizes. $\mathcal{N}_{E}(t_w)$ is expected to be proportional to system size $L$ and therefore, $\mathcal{N}_{E}(t_w)/L$ should follow a master curve as shown in Fig. \ref{coarsening_characteristics}(c).

(ii) The length scale $\mathcal{L}(t_w)$, defined as the separation between two successive $E$ domains, varies as $t_w^{1/2}$ as shown in Fig. \ref{coarsening_characteristics}(d).

(iii) During coarsening, the system has equilibrated within the length scale $\mathcal{L}(t_w)$, which typically contains several $O$ domains. In equilibrium, there are typically $\sqrt{L}$ number of odd domains in a system of size $L$. Therefore, we expect the number of $O$ domains during coarsening to vary as $\sqrt{\mathcal{L}(t_w)}$. Thus, their average number, $\mathcal{N}_{O}(t_w)$, decreases as $t_w^{-1/4}$ as shown in Fig. \ref{coarsening_characteristics}(e) and the average $O$ domain length, $\xi(t_w)$, increases as $t_w^{1/4}$ as shown in Fig. \ref{coarsening_characteristics}(f).

(iv) The decrease of $\mathcal{N}_{O}(t_w)$ shows that some $O$ domains get annihilated during coarsening. To understand how this happens, consider a particular configuration $\Od\Eu\Od$. The number of spins in $\Eu$ decreases through a succession of double spin flips until it reduces to width two, following which $\Eu$ is annihilated. The two $\Od$ domains then join and create a single $\Ed$ domain  as highlighted in the region A in Fig. \ref{coarsening_characteristics}(a).

(v) There are two separate mechanisms by which $E$ domains get annihilated:
\begin{itemize}
	\item
	Consider a configuration $\Od\Eu\Ed$. When $\Eu$ is annihilated, the entire domain becomes $\Od$ as highlighted in the region B in Fig. \ref{coarsening_characteristics}(a).
	\item
	If there are successive $E$ domains, such as $\Ed\Eu\Ed$, annihilation of $\Eu$ creates an $\Ed$ domain. We find this second mechanism becomes rarer as coarsening proceeds.
\end{itemize}

(vi) Consider a configuration where an $E$ domain is surrounded by several $O$ domains on either side, for instance $\Od\Ou\Od\Eu\Od\Ou\Od$. Through the mechanism (iv) above, $O$ domains will continue getting annihilated till the configuration meets an $E$ domain on one side and an $O$ domain on the other side. In this case, annihilation of the central $E$ domain creates a broad $E$ domain that becomes narrower with time and eventually gets annihilated creating another broad $E$ domain:
\begin{align}
\Ou\Od &\Ou\Od\Eu\Od\Ou\Od\Eu \Rightarrow \Ou\Od\Ou\Ed\Ou\Od\Eu \nonumber\\ 
&\Rightarrow \Ou\Od\Eu\Od\Eu \Rightarrow \Ou\Ed\Eu \Rightarrow \Ou \nonumber 
\end{align}
Such a process is highlighted by the region C in Fig. \ref{coarsening_characteristics}(a).

(vii) To understand the $t_w$ dependence of the typical $O$ domain length $\xi(t_w)$, we follow the distribution $P(\xi(t_w))$. We find that $P(\xi(t_w))\sim \exp(-\xi^{1/2})$ as shown in Fig. \ref{coarsening_characteristics}(g). Considering $\xi\sim t_w^\alpha$, we have
\begin{equation}
P(\xi(t_w))\sim \f{1}{t_w^\alpha}e^{-(\xi/t_w^\alpha)^{1/2}}.
\end{equation}
Plotting $t_w^\alpha P(\xi(t_w)) $ as function of $\xi/t_w^\alpha$, we obtain excellent data collapse for $\alpha=1/4$ for different $t_w$ as shown in Fig. \ref{coarsening_characteristics}(h). This confirms our earlier expectation that $\xi\sim t_w^{1/4}$.

\section{Time evolution of the system when initial state is prepared with zero sublattice magnetization $\M$}
\label{slmzero}
As discussed in Sec. \ref{domainwalls}, the number of domain walls $N$ equals the sublattice magnetization $\M$. This implies that when a system with an initial state that has $\M=0$ is quenched to $T=0$, it evolves to a steady state with zero domain walls, that is a state with all spins $\uparrow$ or $\downarrow$. 
To test this result, we have prepared random initial states with $M_{even}=M_{odd}$ with different values of $M_{even}$ for a range of system sizes $L$. A typical time evolution for a quench to $T=0$ from such an initial sate is shown in Fig. \ref{slmzeroevolution} where black refers to $\downarrow$ and yellow refers to $\uparrow$. We find that the system goes to a state with all spins either $\uparrow$ or $\downarrow$.



%

\end{document}